\documentclass[conference, 10pt]{IEEEtran}
\usepackage{booktabs}
\usepackage{graphicx}
\usepackage{balance}
\usepackage{xcolor}
\usepackage{algorithm}
\usepackage{algpseudocode}
\usepackage{enumitem}
\usepackage{subfigure}
\usepackage{url}

\def\note#1{}

\begin{document}

\title{A Look at Communication-Intensive \\Performance in Julia}

\author{\IEEEauthorblockN{Amal Rizvi}
\IEEEauthorblockA{\textit{Department of Computer Science}\\
\textit{Illinois Institute of Technology}\\
Chicago, IL, USA\\
arizvi1@hawk.iit.edu}
\and 
\IEEEauthorblockN{Kyle C. Hale}
\IEEEauthorblockA{\textit{Department of Computer Science}\\
\textit{Illinois Institute of Technology}\\
Chicago, IL, USA\\
khale@cs.iit.edu}
}

\maketitle              % typeset the header of the contribution

\begin{abstract}

The Julia programming language continues to gain popularity both for its
potential for programmer productivity and for its impressive performance on
scientific code. It thus holds potential for large-scale HPC, but we have
not yet seen this potential fully realized. While Julia certainly has the
machinery to run at scale, and while others have done so for embarrassingly
parallel workloads, we have yet to see an analysis of Julia's performance
on communication-intensive codes that are so common in the HPC domain. In
this paper we investigate Julia's performance in this light, first with
a suite of microbenchmarks within and without the node, and then using the
first Julia port of a standard, HPC benchmarking code, high-performance
conjugate gradient (HPCG). We show that if programmers properly balance the
computation to communication ratio, Julia can actually outperform C/MPI in
a cluster computing environment.

\end{abstract}

\section{Introduction}
\label{sec:intro}

The HPC community has been slow to adopt high-level, managed
languages---increasingly popular in industry---due to their apparent
performance overheads and due to the fact that rarely are these languages
designed with the parallelism or behavior of large-scale scientific workloads
in mind. However, such languages hold promise because they can relinquish the
programmer from the burden of low-level details of the system like pointers,
memory management, and type specification. This frees domain scientists from
having to be systems programmers  and allows them to instead focus on their
application logic, thus reducing the ``time to science.''

The {\em de facto} standard for programming large-scale machines involves some
combination of MPI paired with C/C++/Fortran for inter-node parallelism, and
runtime support such as OpenMP~\cite{DAGUM:1998:OPENMP},
pthreads~\cite{BUTTLAR:1996:PTHREADS}, Cilk~\cite{BLUMOFE:1996:CILK}, or
Qthreads~\cite{WHEELER:2008:QTHREADS} for handling intra-node parallelism.
More recently, applications leverage heterogeneous hardware by using
accelerator frameworks like OpenCL~\cite{STONE:2010:OPENCL},
OpenACC~\cite{OPENACC} or CUDA~\cite{CUDA_PROG_GUIDE}.  For domain experts,
this makes the problem worse, since not only must they worry about the ``how''
of parallelism across nodes, they must also deal with accelerator offloading,
synchronization, virtual memory, and other low-level details.

Ideally, the programmer would describe their problem in a high-level language
and the underlying machinery of the language would take care of mapping the problem
onto parallel and heterogeneous hardware. While we are not there yet, the
community {\em has} made strides in language design, compiler technology, and
parallel programming frameworks. Recent languages like
Regent~\cite{SLAUGHTER:2015:REGENT} and more established languages like
Chapel~\cite{CHAMBERLAIN:2007:CHAPEL}, HPF~\cite{MERLIN:1995:HPF},
UPC~\cite{CCS:1999:UPC}, X10~\cite{CHARLES:2005:X10}, and
Erlang~\cite{JOHANSSON:2000:HPE} all have made significant contributions in
bringing productive programming to the HPC world, yet they have not seen
widespread adoption.  For a language to flourish in the HPC community, it must
be sophisticated (sporting a rich set of built-in functionality for scientific
programming), intuitive, and mature. Most importantly, however, {\em it must
perform well} for the increasingly broad set of large-scale applications that
the scientific community employs. 

The Julia language\footnote{\url{https://julialang.org}} is an interesting
point on this spectrum not only because it takes inspiration from other popular
managed languages like Python and Ruby, but also because it was designed by
domain experts with scientific and technical computing in mind.  It thus has
significant appeal for programmers brought up on Fortran and MATLAB. Unlike
those languages however, Julia has targeted scalable parallelism from the
start. Built atop the LLVM compiler framework, and sporting sophisticated
features like dynamic dispatch, high-performance math libraries, and JIT
compilation, Julia's single-threaded performance for scientific problems
approaches (or even surpasses) its low-level counterparts, C and Fortran. It
also features intuitive parallel programming abstractions, and others have
recently demonstrated its potential for petascale parallelism on a top
supercomputer~\cite{Regier2015CelesteVI}.

However, a large class of HPC applications involve significant inter-node
communication to achieve parallelism. A classic example is stencil computations
following the bulk-synchronous parallel (BSP) model, where in order to solve,
e.g. a partial differential equation (PDE) describing some dynamical system,
the domain is discretized and evolved in time using finite differences. While
the programmer must carefully balance the communication/computation ratio with
these methods, communication latencies due to language, library, system
software, and network interfaces can make or break performance.

As far as we are aware, the Julia language has not yet been evaluated in this
respect.  Thus, in this paper, we seek to shed light on Julia's performance
for communication-intensive applications, both within and without the node. We
investigate Julia using a range of single-node and multi-node microbenchmarks,
as well as a port of a standard HPC benchmarking code encapsulating the above
behavior, high-performance conjugate gradient (HPCG).

We make the following contributions:
\begin{itemize}[nosep]
    \item We provide a series of single-node and multi-node microbenchmarks for
        Julia. 
    \item We evaluate the performance of Julia's parallel programming abstractions,
        both on a single node and in a distributed environment.
    \item We present the first port of a standard, large-scale, communication-intensive code (HPCG) to Julia. 
	\item We show through benchmarking that with a reasonable computation-to-communication
        ratio, Julia can actually outperform C in distributed message passing scenarios,
        demonstrating its potential for communication-intensive applications.
\end{itemize}

All code for our benchmarks and our experiments, as well as our experimental
data will be made freely available upon acceptance. 
The remainder of this paper is organized as follows:
Section~\ref{sec:background} gives an overview of the Julia language and its
focus on scientific computing and parallelism. Section~\ref{sec:eval} presents the evaluation of Julia
for high-performance parallelism within the node. In Section~\ref{sec:distributed}, 
we analyze Julia's performance in a distributed environment. We discuss 
our results in Section~\ref{sec:discuss}, outline related prior art 
in Section~\ref{sec:related} and conclude in Section~\ref{sec:conc}.

\section{Background}
\label{sec:background}

In general, to be effective for efficient parallelism, a high-level language
will have high-performance primitives or constructs which support:

\begin{itemize}
	\item Mapping execution contexts to hardware precisely (e.g. NUMA and processor affinity)
	\item Hardware heterogeneity
	\item Low-latency communication between execution contexts (threads/tasks/processes)
	\item Low-latency creation and scheduling of tasks.
	\item Efficient exploitation of parallelism (e.g. parallel \verb.map().s or parallel loop constructs)
\end{itemize}

Much of the underlying machinery for the above will be hidden from the
programmer. 

Ideally the performance of the language will be close to the hardware's
capabilities, as is the case with lower-level languages. Some (including
Julia's designers) have made the argument that even if this is not true for the
HLL, the benefit to programmer productivity and application development time
should be considered as a counterbalance to the performance of the language,
especially since development time can take several years for large HPC codes.
This perspective may hold even more weight as new generations of domain experts
and programmers are being taught with high-level languages.  Unfortunately, this
aspect of a language is more challenging to quantitatively measure in practice.
If the HLL's performance {\em can} match that of its lower-level counterpart,
all the better, since we get the best of both worlds. We seek to find out if
that is indeed the case for Julia. 

Below we will outline the parallel programming primitives that Julia provides
and discuss some salient choices by the language designers.

%Some HPC application developers like machine-learning researchers ~\cite{Dinari2019JuliaMCMC} 
%hold the perspective that if performance is close enough to that of C/Fortran 
%if not better then these abstraction and ease of use 
%features are enough to balance out the shortcomings of performance. 
%It can be argued that a lesser cost of application development both monetary and
%in terms of time taken to write application in HPC
%amortizes the cost incurred in performance that is incurred in using a
%high level language. In our opinion this does hold true, but only if the
%the overheads are sustainable and the features, ease of use and parallel
%construct support is robust.

\subsection{Overview of Julia}
\label{sec:julia}

Julia~\cite{BEZANSON:2017:JULIA} is a high-level programming language designed
for numerical computing. Our main interest in Julia stems from its promising
single-threaded performance (which the Julia developers have shown can approach
that of C and Fortran), its native parallel constructs, its extensibility, and
its rich set of user-developed packages. Particularly its performance relative
to other high-level languages, which is in part due to its usage of LLVM's JIT
compilation framework, indicates that it might be a suitable candidate for
communication-intensive HPC applications as well as for building parallel
runtime systems which support even higher-level (possibly domain-specific)
parallel languages. Some other noteworthy features of the Julia language include
multiple dispatch, dynamic types, meta-programming features, light-weight user
threads, easy support for native code invocation, and packages for distributed,
parallel computing. 

Julia is already being used to solve a wide array of problems in technical
computing.  For example, JuliaFEM~\cite{Rapo_Aho_Frondelius_2017} provides
a concise programming framework for solving large-scale problems using finite
element methods. Others have explored using high-level frameworks for GPU
computing with Julia~\cite{2016arXiv160400341Y}. Chen et al. investigated
using polymorphic code features in Julia applied to parallel and distributed
computing~\cite{Chen2014PrefixPolymorphism}. It has also been shown to work
well for classic scientific problems, such as convex
optimization~\cite{Udell2014ConvexJulia}, PDE
solving~\cite{Heitzinger2014JuliaPDEs}, and discrete event-driven
simulation~\cite{Kourzanov2018JuliaLiveSim}. Data scientists and AI experts are
increasingly employing Julia with success as well~\cite{Dinari2019JuliaMCMC,
Regier2015CelesteVI}.

\subsection{Parallelism in Julia}
\label{sec:parallelism}

Julia has an extensive array of features and packages that support parallel and
distributed computing. Julia natively provides an abstraction for task
parallelism called {\em Tasks}, which are essentially lightweight user threads.
Once tasks are defined they are invoked using the \verb.@schedule. macro.

{\em Workers} are an abstraction of processes and \verb.workers(). is
a function that returns an array of all workers deployed locally or
remotely. Processes do not share memory by default but can communicate
via \verb.Channels. or by using \verb.SharedArrays., discussed later
in this section.  Multiple processes can be created locally and on other
nodes by calling \verb$@addprocs$.

{\em Threads} are Julia's version of kernel threads. Others have evaluated the
native threading support on multi-core
machines~\cite{Knopp2014JuliaMultiThreading}. With these threads, Julia allows
the user to wrap a value in \verb|Threads.Atomic| to avoid race conditions
while accessing or modifying shared state. Similar to OpenMP, one can indicate
that a code region is to be run on multiple threads using the \verb.@threads.
macro. A barrier is implicitly placed at the end of a \verb.@threads. block and
the code in that block will run on the number of threads specified by the
\verb.JULIA_NUM_THREADS. environment variable, again similar to standard OpenMP
programming practice.  Additionally, Julia's threading package includes a set
of native atomic operations (usually provided by compiler built-ins in
lower-level languages) via LLVM, in addition to higher-level synchronization
primitives like locks, mutexes, and semaphores.

Communication between Julia processes is, by default, one-sided. It is
accomplished using the \verb.RemoteReference. and \verb.RemoteCall.
abstractions.  A \verb.RemoteReference. is made for exchanging information
while \verb.RemoteCall. is used to request the remote execution of a function.
\verb.RemoteCall. returns a \verb.Future. object asynchronously---essentially
a \verb.RemoteReference.---and the result of the function can be obtained by
calling \verb.fetch. on the \verb.Future. object.  The \verb.@spawn. and
\verb.@spawnat.  macros are used to send work to an arbitrary and particular
process, respectively. Julia also provides an \verb.@everywhere. macro which
specifies that its wrapped expression shall run on all available workers.

Fine-grained parallelism is provided by \verb.@distributed. (for indicating
that loop iterations run in parallel) and \verb.pmap. (parallel mapping of
a function over a collection). The Julia documentation points out that if the
number of operations are large in number but not too complex (e.g. using synchronization), 
\verb.@distributed. can be
used to perform them.  On the contrary, if the operations are complex and not
numerous then use of \verb.pmap. is recommended. 

\verb|@async| is a feature that can be used in scenarios where one wants to define
work but is not ready to have them executed right away. \verb.@async. will
launch tasks locally and can be sent to other processes or workers using
\verb.remotecall_fetch().. 
Julia has macros for specifying operations as \verb.@sync. and \verb.@async. as
per need.  \verb.@sync. can be employed on an otherwise \verb.@async. block and
vice versa. This, however,  does not mean that Julia will block a task that uses another \verb.@async. 
task. \verb.@async.  can only block a synchronized function.

%\note{Can I do this at an arbitrary granularity? What about dependent statements? Will
%they block even if I use Async?}

\verb.Channels. and \verb.RemoteChannels. allow the programmer to construct
producer/consumer queues between local and remote processes, respectively.
This is useful for transfer of primitive data across tasks. For more complex
data sharing like an array, \verb.SharedArray. and \verb.DistributedArray. are
used.  A \verb.SharedArray. can be accessed by all processes in parallel, with
synchronization provided by the runtime.  A \verb.DistributedArray.  is similar
in concept but has the advantage of decomposing the array by distributing chunks
across local and remote processes.

Julia supports this set of parallel and distributed operations and
structures via the external packages \verb$Distributed.jl$ and \verb$Threads.jl$. 
Importantly, Julia also sports a package to provide MPI bindings, which we will
leverage in Section~\ref{sec:distributed}.

While other features of Julia are beyond the scope of this paper, we refer
interested readers to other texts on the
language~\cite{DBLP:journals/corr/abs-1209-5145, Cabutto2018Julia,
Bezanson2018Julia, Zappa2018JuliaSubtyping} and its design choices, as well as
the official Julia documentation\footnote{\url{https://docs.julialang.org}}.

\section{Parallelism within the node} 
\label{sec:eval}

In this section we present a series of experiments evaluating the performance
of Julia's runtime native parallelism primitives within a node compared to 
low-level implementations in C.  Our goal is to get a sense of what kinds of barriers to
parallel performance an HPC programmer might encounter on a compute node when
using Julia.  

\subsection{Experimental Setup}
\label{sec:exp-setup}

All experiments were carried out on our in-house, 14-node cluster called {\em
mystic}.  Each node comprises a dual-socket (2 NUMA nodes), 16-core (2 hardware
threads per physical core) Intel Xeon Silver SP 4108 (Skylake) CPU clocked at
1.80GHz and equipped with 64GB of 2.66 GHz DDR4 SDRAM.  This machine runs
  Ubuntu 18.04 LTS with a stock Linux kernel version 4.15. We configured the
  BIOS to disable DVFS features. 

For the Julia experiments, we use Julia version 1.3.1. All C/C++ benchmarks
were compiled with the highest optimization level (\verb.O3.) using the stock
Ubuntu version of gcc (7.5). Unless otherwise noted, all experiments were
conducted using 100 trials, with an additional 10 trials initially thrown out
to mitigate cold start effects. Timing was conducted using Julia's
native \verb.time_ns().  function and \verb.clock_gettime(). in C using the
\verb.CLOCK_REALTIME. time source. The file system configuration has no influence
on our experiments. 

Many high-level languages support the notion of a {\em task}, a schedulable
unit of work which may or may not have the ability to modify state. Tasks can
essentially be thought of as glorified functions, but they can be moved around
the system.  Tasks may be mapped to threads or processes, but ultimately they
should be very light-weight. The more overhead that the language introduces
when creating and managing tasks, the coarser the units of work must be to
amortize these costs, and the amount of parallelism available to the programmer
is reduced. Julia has support for such tasks (with the \verb.Task.
abstraction), which they describe alternatively as light-weight coroutines or
{\em green threads}. These are essentially user-level threads of which the
underlying OS is unaware. Our first experiment measures the latency of creating
these tasks. The baseline we compare this against is the time taken to create a pthread
(using \verb.pthread_create().). Since pthreads map directly to kernel threads
(visible to the OS scheduler), we expect that the Julia tasks should be much
cheaper to create. A comparison to an equivalent low-level user-threading package
(e.g. Qthreads~\cite{WHEELER:2008:QTHREADS} or OpenMP tasks) would be ideal
here, but we note that this is just to sanity check the task performance of
Julia. We leave such a comparison for future work.

To make the comparison as fair as possible, in Julia we measure the time to
create a task using the \verb.Task(). function followed by a call to
\verb.schedule(). which places the created task on the task scheduler's run
queue (which is local to the current Julia process on a single core). Note that
unlike most \verb.schedule(). functions, this Julia version does not
immediately yield to the scheduler.

\begin{figure} 
    \centering 
    \subfigure[]{
        \includegraphics[width=0.35\columnwidth]{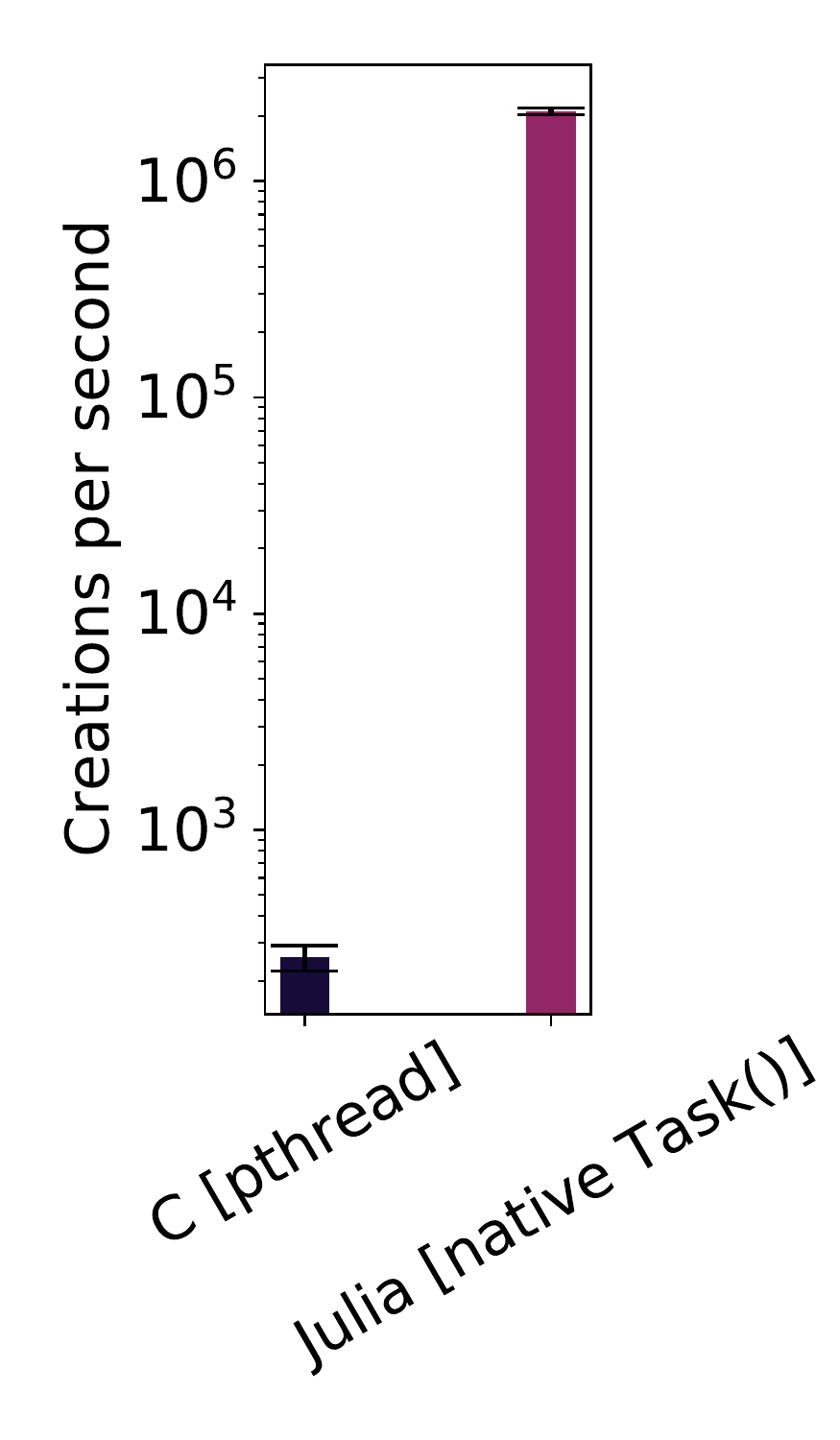}
        \label{fig:task-create} 
    } \subfigure[]{
        \includegraphics[width=0.35\columnwidth]{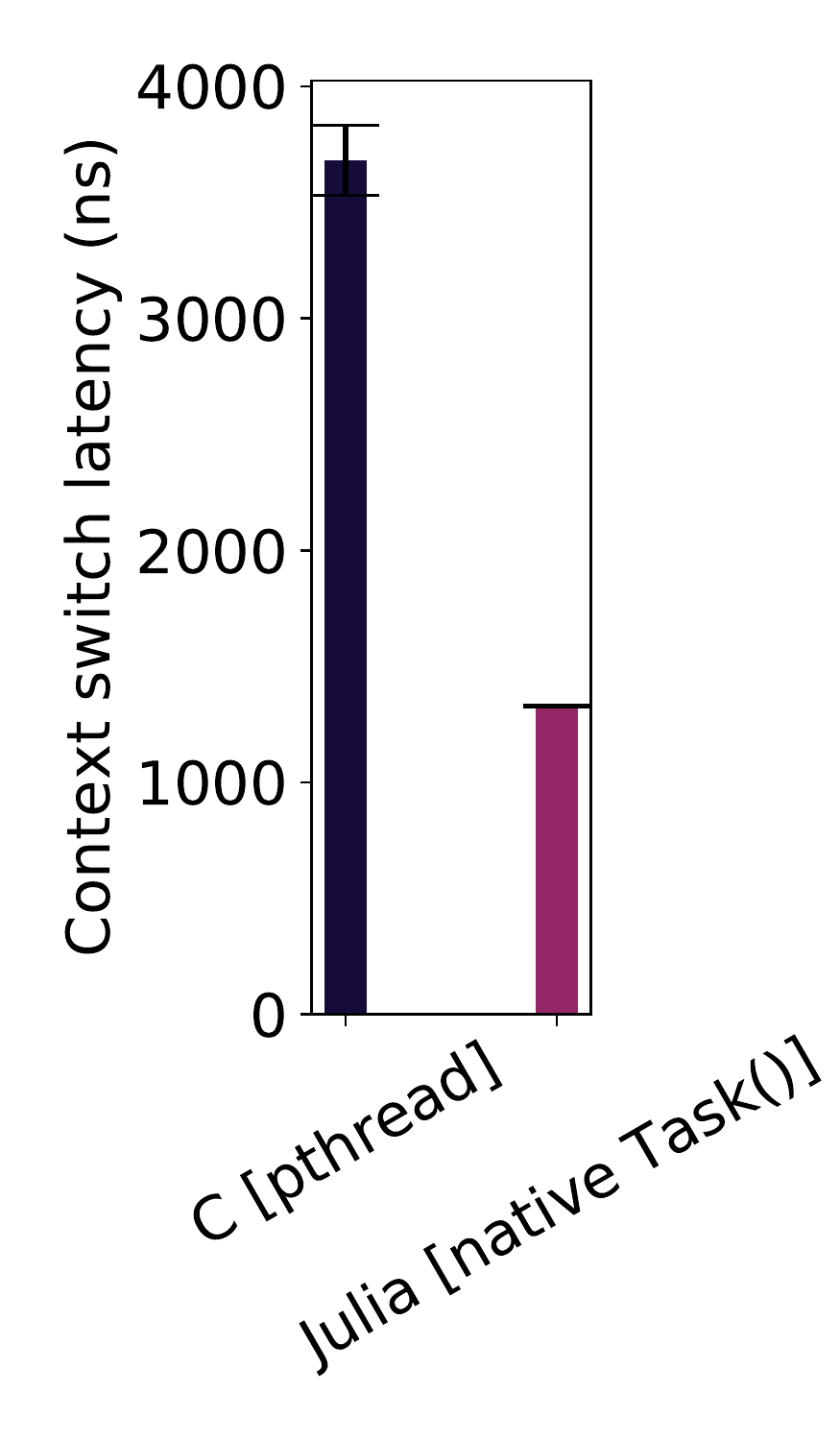}
        \label{fig:ctx-switch} 
    } 
    \caption{(a) Comparison of task creation latencies for Julia's native tasks and
    kernel threads. (b) Context
    switching latencies for Julia tasks and pthreads.} 
    \label{fig:tasks}
        
\end{figure}

Figure~\ref{fig:task-create} shows the results (note the log-scale on the y axis). Creation of a pthread (kernel thread)
costs roughly 20$\mu$s while Julia task creation and the subsequent schedule call
costs only about 400ns. This is consistent with the creation throughput shown in the figure, and is in line with our expectations. 
However, tasks
on a remote core must be scheduled on a remote worker process, which must be
explicitly created with the \verb.addprocs(). routine. While this routine takes
roughly 400ms, one need only call it once, at runtime startup. 
% (in spawn/fetch)

% NOTE: addprocs avg is 4.1586151569e8 NOTE: julia task avg is 398.7409  

Often in HPC scenarios (i.e. many-task computing), there are many more tasks
than there are hardware threads available in the system. This allows a high
degree of concurrency when tasks perform blocking operations. It is important
that tasks be light-weight to maximize this concurrency. Here we measure the
context switching overhead of Julia tasks by creating two tasks in the same
worker process, having them cooperatively \verb.yield(). to each other a fixed
number of times, and approximating the latency from the number of yields in the
recorded time period.  The same caveat as above applies to comparing kernel
threads with user threads. For our baseline we use the same strategy with two
pthreads pinned to the same core.

Figure~\ref{fig:ctx-switch} shows that Julia task context switching is roughly
one order of magnitude faster than for pthreads. Again, this is to be expected
since the kernel is not involved in a Julia context switch unless a task yields
and there is nothing else to run.

%These results tell us that Julia should encounter no significant issues
%in many-task computing scenarios.

\begin{figure}
	\centering
    \includegraphics[width=0.9\columnwidth]{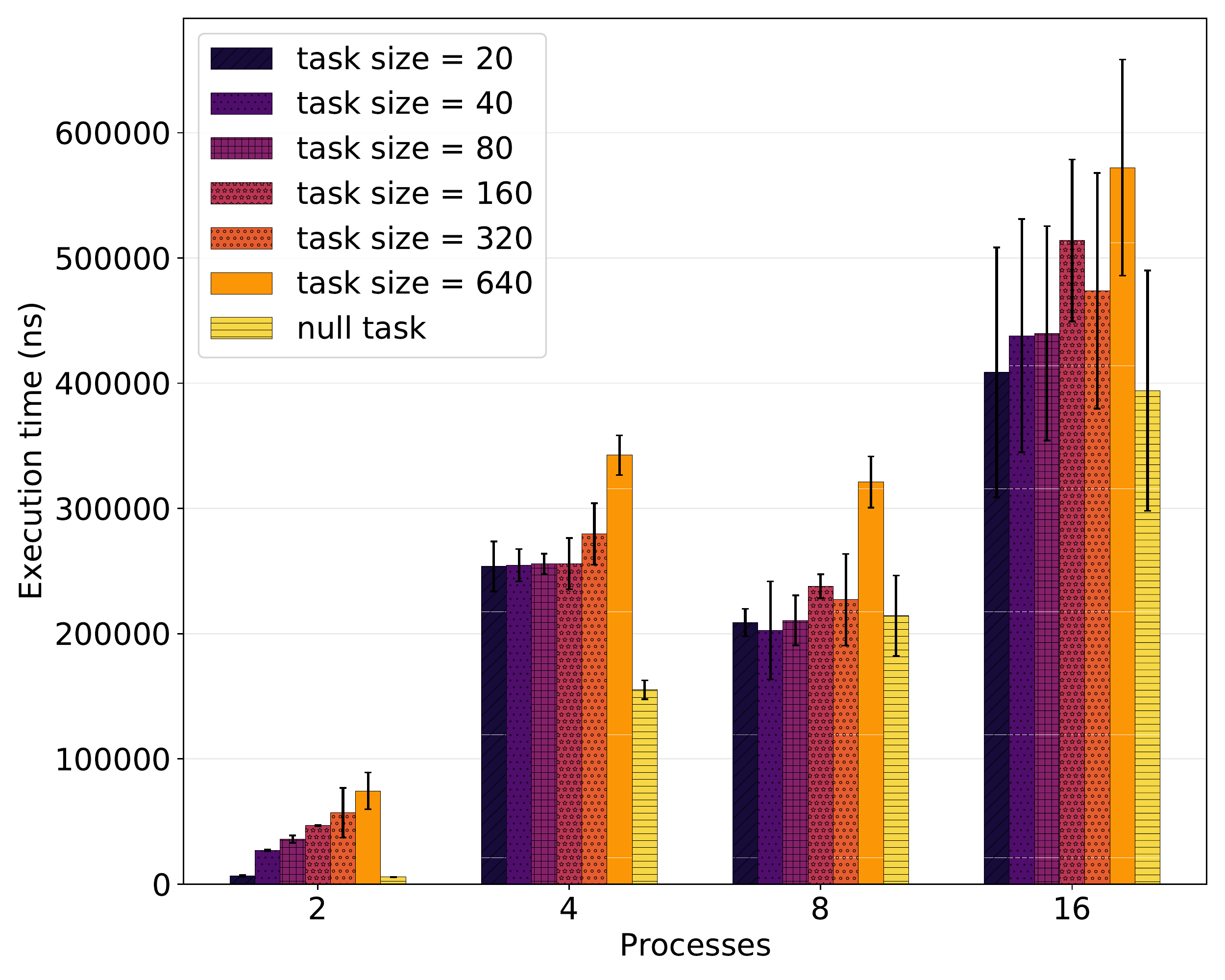} 
    \caption{Mean latencies for Julia's one-sided future-based
    parallelism (spawn and fetch).}
    \label{fig:spawn_fetch}
\end{figure}

\begin{figure}
	\centering
    \includegraphics[width=0.7\columnwidth]{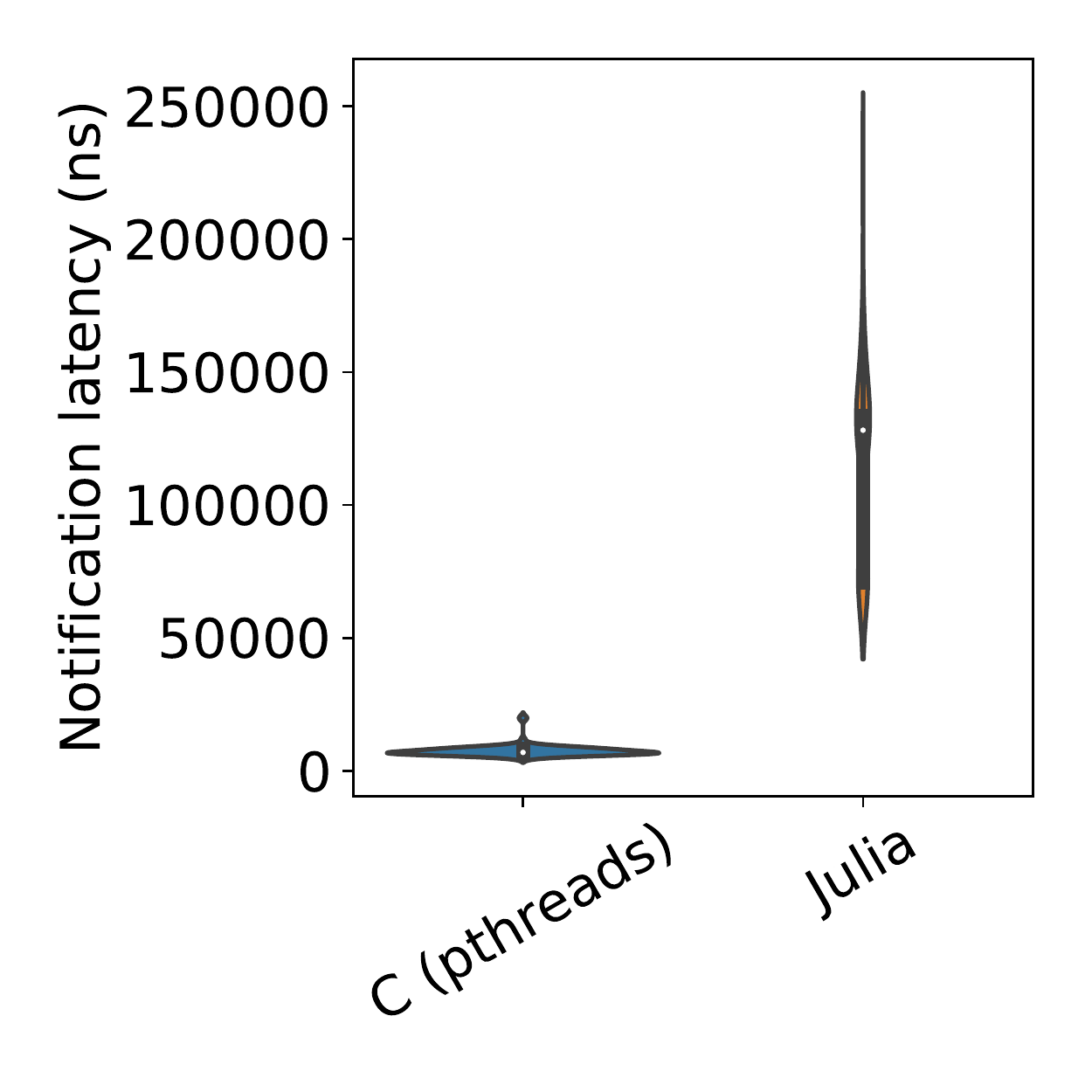}
    \caption{Event notification latency for Julia and pthread condition variables.}
    \label{fig:condvars}
\end{figure}

For the next experiment, we attempt to get an estimate of the overheads of
task operations between Julia workers.
Julia's native parallel processing features are
based on one-sided task-launch operations (futures).
When a future is created (using \verb.spawn().), its corresponding object is
immediately returned to the caller, without blocking. The work associated with the future may
complete immediately, or at some later point in time. The result is retrieved by
an explicit call to (\verb.fetch().) the future. This call
{\em will}
block if the execution has yet to take place.
Here we aim to investigate how the overheads of task creation relate to
the amount of work performed for fine-grained tasks.
For this experiment we created two benchmarks:
a tunable benchmark which returns the first $n$ Fibonacci numbers sequentially
and a baseline benchmark which creates a null task that performs no actual
work at all. The null task thus approximates the overheads of remote
task creation. 

Figure~\ref{fig:spawn_fetch} depicts the results of spawning a remote task (on
any available worker) for various task granularities (where again, the
granularity is determined by the \verb.fib(). argument). These are compared
against the null task as we vary the number of Julia workers present in the
system. We cap our experiment at
16 worker processes since the system has 16 physical cores. Note that here we
are only spawning the task on {\em a single} remote worker, so the latency
should remain the same, independent of the number of workers. However, looking
at the figure this is not what we see.  Instead, it seems that Julia's task
scheduling across workers scales roughly linearly with the number of
workers. While we make sure to create workers (\verb.@addprocs.) {\em
before} the experimental run, we suspected that Julia may be lazily creating
workers on the spawn invocation. However, since we are throwing out the cold
start trials, we believe this not to be the case.  We found this effect
surprising, and we have not yet determined its cause.  These results were
also confirmed by other comparisons~\cite{Huang2019JuliaDirichlet} between
inter-process runs and inter-node runs. Note that while the error bars (depicting
one $\sigma$ above and below the mean) are relatively small, we did see several
significantly large outliers in this experiment in Julia's case. The measurements
shown here are {\em after} outlier removal using Tukey's method with a constant of 1.5\footnote{
    That is, all measurements not on the interval $[x_{25\%} - 1.5IRQ, x_{75\%} + 1.5IQR]$ have
    been removed.}.

%% EVENTS/COMMUNICATION

Often high-performance applications will need low-latency communication. This
is needed to support explicit communication between tasks (e.g. using explicit
messaging), implicit communication based on access to shared data, and event
notification (e.g. in a parallel runtime system).

Event notification can signify conditions such as the creation of
a task, the availability of data, an error condition, or the completion of
a task. These notifications are
particularly important for processes communicating via event channels.
One of the most common software event notification
mechanisms is a {\em condition variable}. A condition variable is essentially
a queue protected by a mutex, and its interface comprises three functions:
\verb.wait()., \verb.signal()., and \verb.broadcast().. A thread/task calls 
\verb.wait(). to block execution until some {\em condition} becomes true (e.g.,
a unit of work is ready to execute, a message delivery has completed). The
thread will then be descheduled and put on the condition variable's queue 
until the condition becomes true, thus yielding the processor to another unit of
execution. Some
other thread will then call \verb.signal(). to wake one waiting task and remove
it from the queue, or call \verb.broadcast(). to wake all waiting tasks.
A common model in runtimes/HPC applications is to have worker threads/tasks spread across cores,
where they wait on work to do with condition variables. Unfortunately, Julia's
\verb.Condition(). construct only applies to \verb.Tasks. on the same worker process,
so there is no way to use them for cross-core notifications. To approximate
this behavior, we used Julia's \verb.RemoteChannel. primitive, which is essentially a
producer/consumer queue that can be referenced by \verb.Task.s running on
different Julia worker processes. We create an unbuffered (0-element)
\verb.RemoteChannel. and issue a \verb.take(). operation to simulate a condition
variable \verb.wait()., and have a \verb.Task. on a worker process running on a
remote core issue a \verb.put(). to simulate a condition variable
\verb.signal().. 
We compare against a baseline implementation written in pthreads, which creates
a remote thread whose job is to signal a condition that the local thread is
waiting on (using \verb.pthread_cond_signal().). The remote thread takes a time
stamp after signaling, and the local thread records a time stamp after waking
up. We report the latency as the difference between these two timestamps. 
The violin plot in Figure~\ref{fig:condvars} shows the results.
We see a median latency of roughly 14$\mu$s for the Julia case, and a few
 microseconds for notification using the pthreads condition variable. The Julia
 version also shows a significantly higher variance in latency. While the
performance gap is quite large, one should realize that the Julia notification
is occurring across worker processes (OS processes) in separate address spaces,
whereas in the pthreads, case this is happening between kernel threads sharing the
same address space. While the notification latency for Julia is already fairly
small, we expect it will shrink once proper threading support (with the
attendant condition variable implementation) is mature.

% CHANNELS
\begin{figure}
	\centering
    \includegraphics[width=0.7\columnwidth]{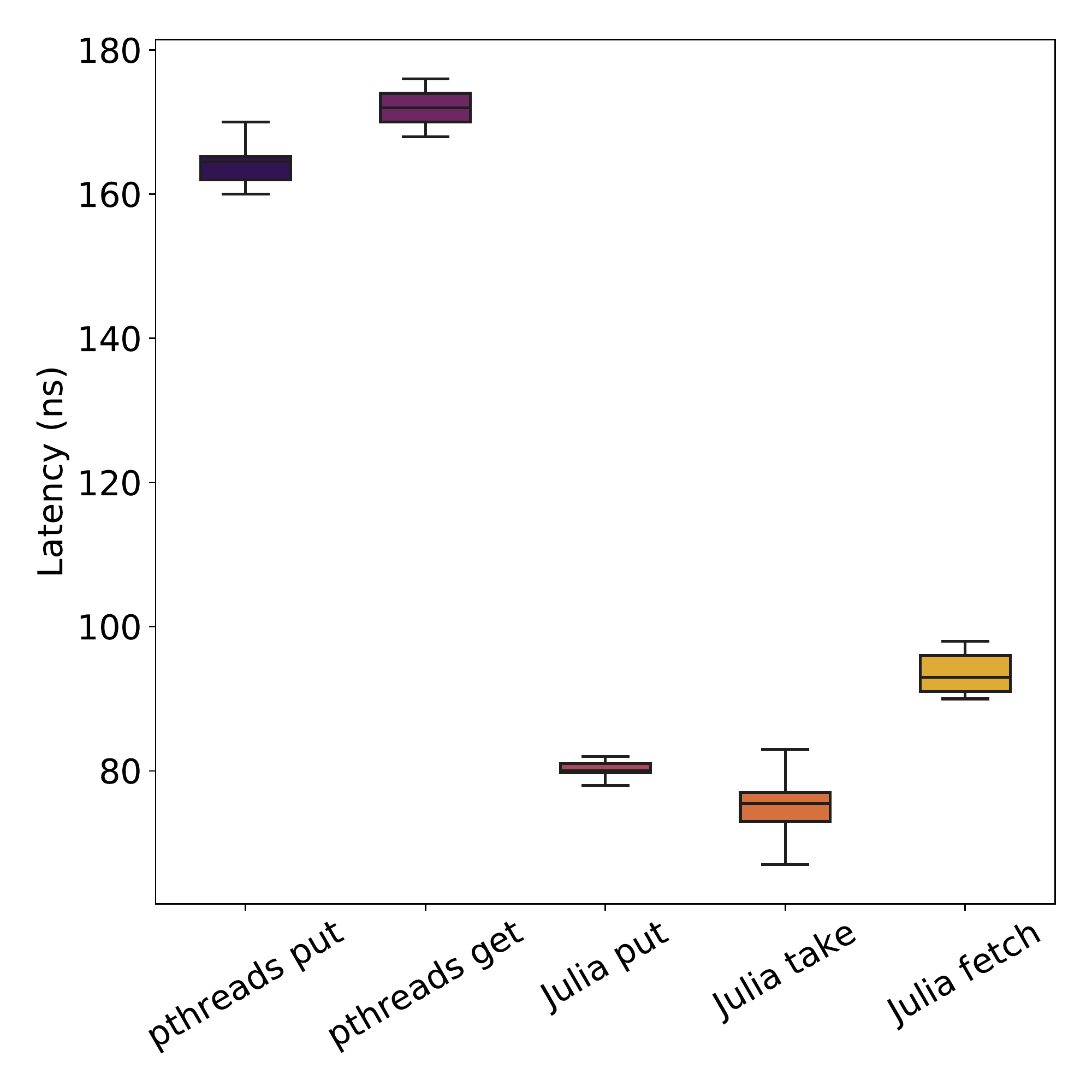} 

	\caption{Latency of channel operations (put/get/fetch) using Julia's channel construct
    and a simple producer/consumer queue written in C with pthreads.}
    \label{fig:channels}
\end{figure}  

In addition to simulating condition variables, channels can also be used to
implement both one-sided and two-sided communication models.  Julia exposes
three operations on channels: \verb.put(). to place an object on the queue,
\verb.take(). to remove an object from the queue, and \verb.fetch(). to read
the front of the queue (without removing the object). Ideally operations on
channels will just be a read or write protected by a lock or an atomic
operation. We compare Julia's channel against a simple (unoptimized)
producer/consumer queue written in C that is protected by a mutex from the
pthreads library (\verb.pthread_mutex_t.).  Figure~\ref{fig:channels}
shows the results. We see here that Julia's channel operations are all very
light-weight, each taking less than 100ns. The differences here with the
C version are not significant, but the operations that Julia provides out of
the box show promising performance.

% PARALLEL OPS
\begin{figure}
	\centering
    \includegraphics[width=0.7\columnwidth]{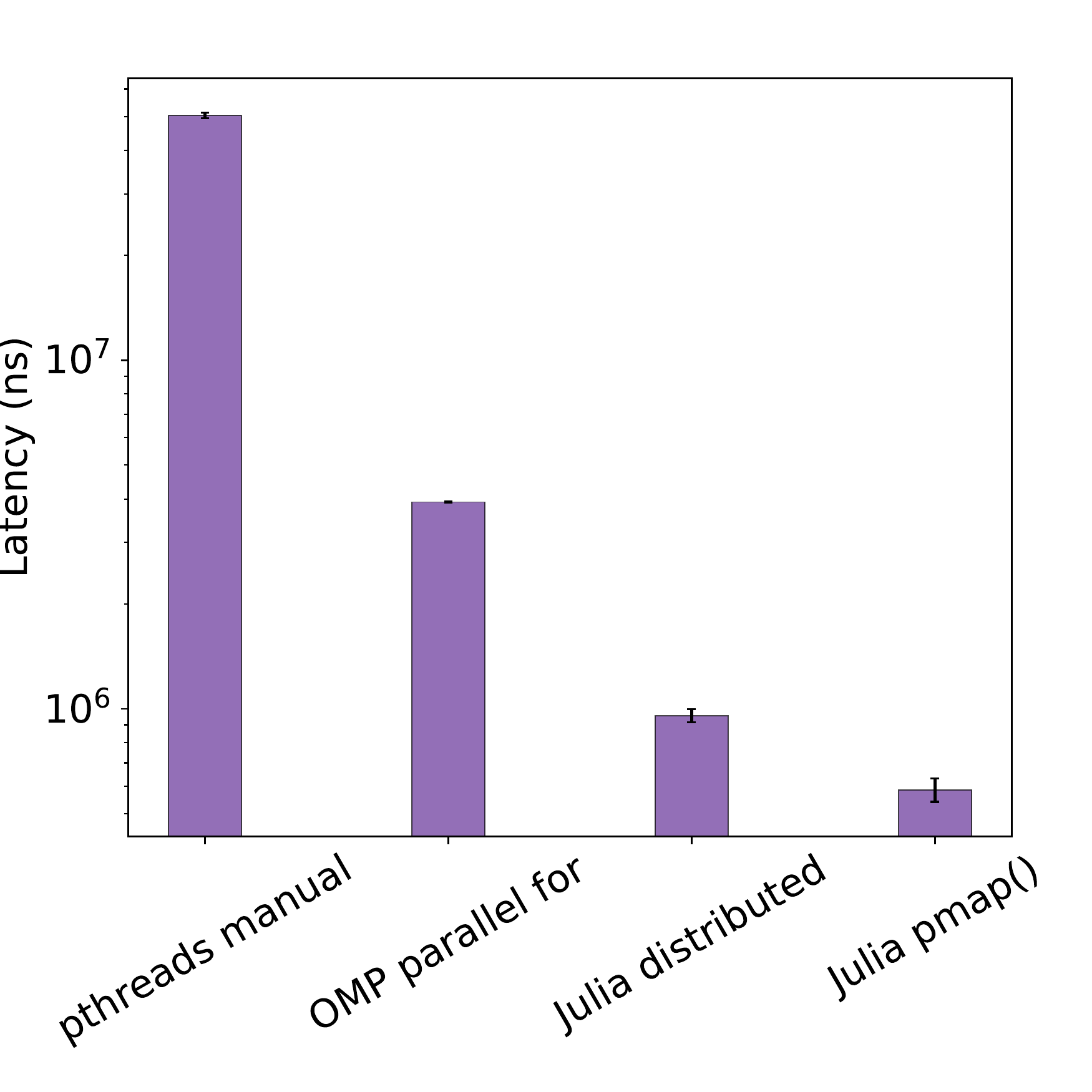} 
	\caption{Mean time to initialize a 50MB
    array across execution contexts.}
    \label{fig:parallel-constructs}
\end{figure}

An important requirement for any HPC-oriented high-level language is that
it exploit parallelism well. To measure how well Julia maps computations automatically
to execution contexts, we distributed the initialization of a 50MB array among worker 
contexts. We use Julia's \verb.pmap(). and \verb.@distributed. features to achieve
this, and compare against an OMP version using the \verb.parallel for. looping
construct and a custom pthreads implementation in C that performs manual decomposition.
Noting the log scale on the y-axis, we see 
in Figure~\ref{fig:parallel-constructs} that the mean time to initialize the 
array using the parallel constructs in Julia is roughly five times smaller.

\section{Distributed Parallelism}
\label{sec:distributed}

To get a realistic picture of how viable Julia is for communication-intensive
HPC applications, we also measure its performance in a multi-node setting.  An
HPC application written in an HLL ideally will have performance close to that
achieved by low-level languages, even when spread across nodes. To do this, we
leverage the \verb$MPI.jl$ package, which provides bindings to the
system-resident MPI library with a set of wrappers that allow overloading of
the typical MPI functions.  For all experiments, we use MPICH v3.3a2. The hardware
configuration was previously described in Section~\ref{sec:exp-setup}. Both
Julia and C/C++ applications use the underlying system MPI library. The 14
nodes in our {\em mystic} cluster are connected in a star topology over 10Gb
Ethernet (via Intel X550 NICs) configured with an MTU of 9000. Timings in the distributed
environment are conducted by having rank 0 (the master node) record timestamps (we plan to 
incorporate measurements from {\em all nodes}).

%\begin{algorithm}
    %\begin{algorithmic}[1]
    %\Function{$do\_comp$}{$n$}
        %\State $i \gets 0$
        %\While{$i < n$}
            %\State $do\_reads(nreads)$
            %\State $do\_writes(nwrites)$
            %\State $do\_floats(nfloats)$
            %\State $i \gets i + 1$
        %\EndWhile
    %\EndFunction

    %\Function{$do\_comms$}{$n$}
        %\State $i \gets 0$
        %\While{$i < n$}
            %\State $send\_forward()$
            %\State $recv\_backward()$
            %\State $barrier()$
            %\State $i \gets i + 1$
        %\EndWhile
    %\EndFunction

    %\State $iter \gets 0$
    %\While{$iter < niters$}
        %\State $do\_comp(ncomp)$
        %\State $do\_comms(ncomm)$
        %\State $iter \gets iter + 1$
    %\EndWhile

    %\caption{Synthetic BSP algorithm.}
    %\label{algo:bsp}
%\end{algorithmic}
%\end{algorithm}

\begin{figure}
    \centering
    \includegraphics[width=0.5\columnwidth]{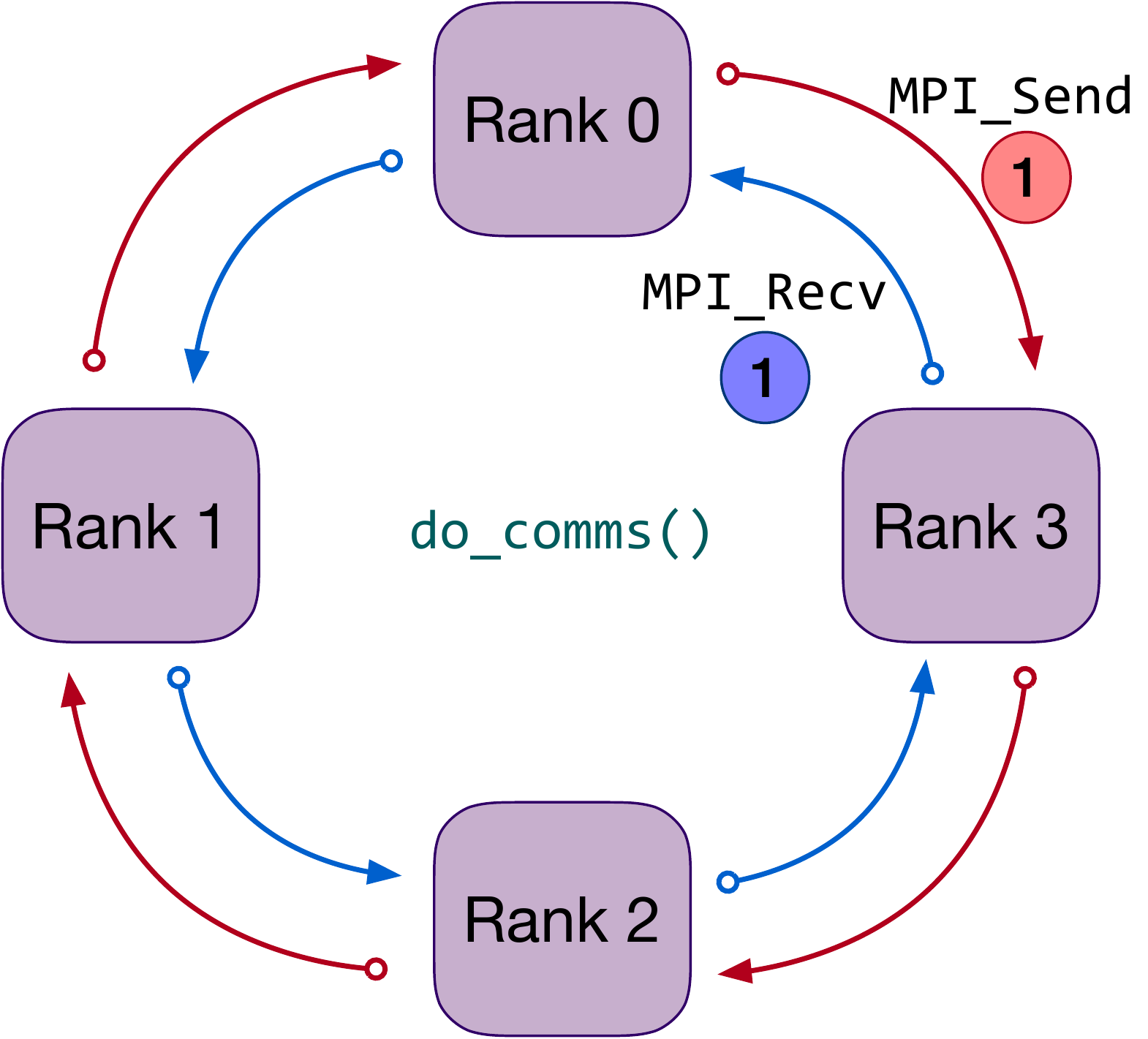}
    \caption{Connectivity between MPI ranks in our BSP benchmark.}
    \label{fig:bsp-topo}
\end{figure}

In our first set of multi-node experiments, we approximate a typical bulk-synchronous
parallel (BSP) code (e.g.  a stencil computation) by creating a synthetic
benchmark which consists of compute (memory reads/writes and floating point
ops) and communication (message send/recv) steps.
Each rank will perform a configurable amount of computation between
communications, wherein computation consists of reads (a series of loads
from an array), writes (a series of stores to an array), and flops (a series of
floating point operations). As shown in the algorithm, in the communication
step, ranks send data to their forward neighbor and then receive data from
their backward neighbor. The ranks thus form a ring topology, as depicted in
Figure~\ref{fig:bsp-topo}. We compare the performance of two custom
implementations of this BSP benchmark in both C and Julia.  We varied the
number of processes (MPI ranks) per node (1, 2, 4, and 8). The amount of work
performed by each rank is kept constant, thus we are technically showing weak
scaling with this experiment. We gather statistics over 100 iterations.

\begin{figure*}
	\centering
    \subfigure[10 communication steps.]{
        \includegraphics[width=0.32\textwidth]{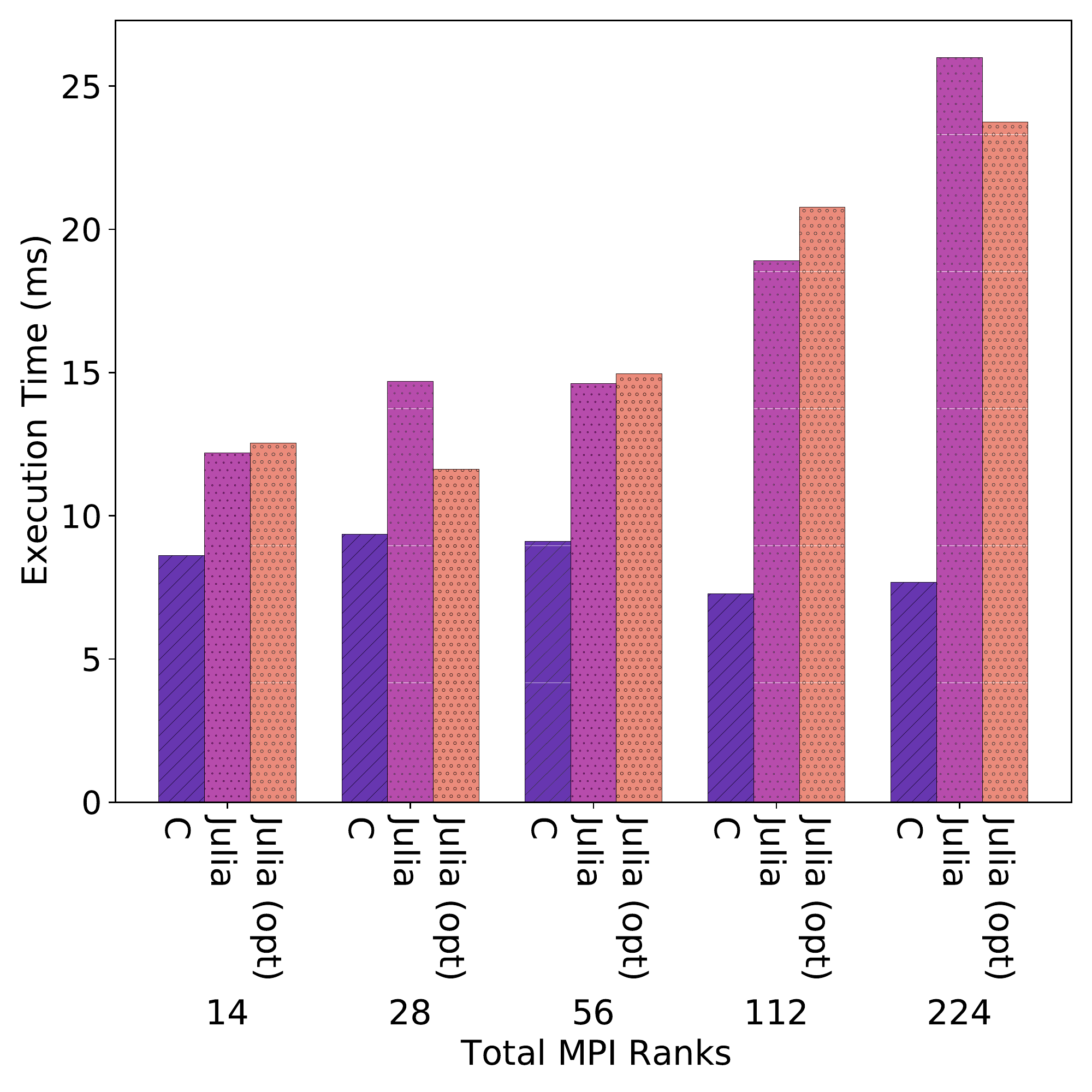}
        \label{fig:bsp-runtimes-10}
    }
    \subfigure[100 communication steps.]{
        \includegraphics[width=0.32\textwidth]{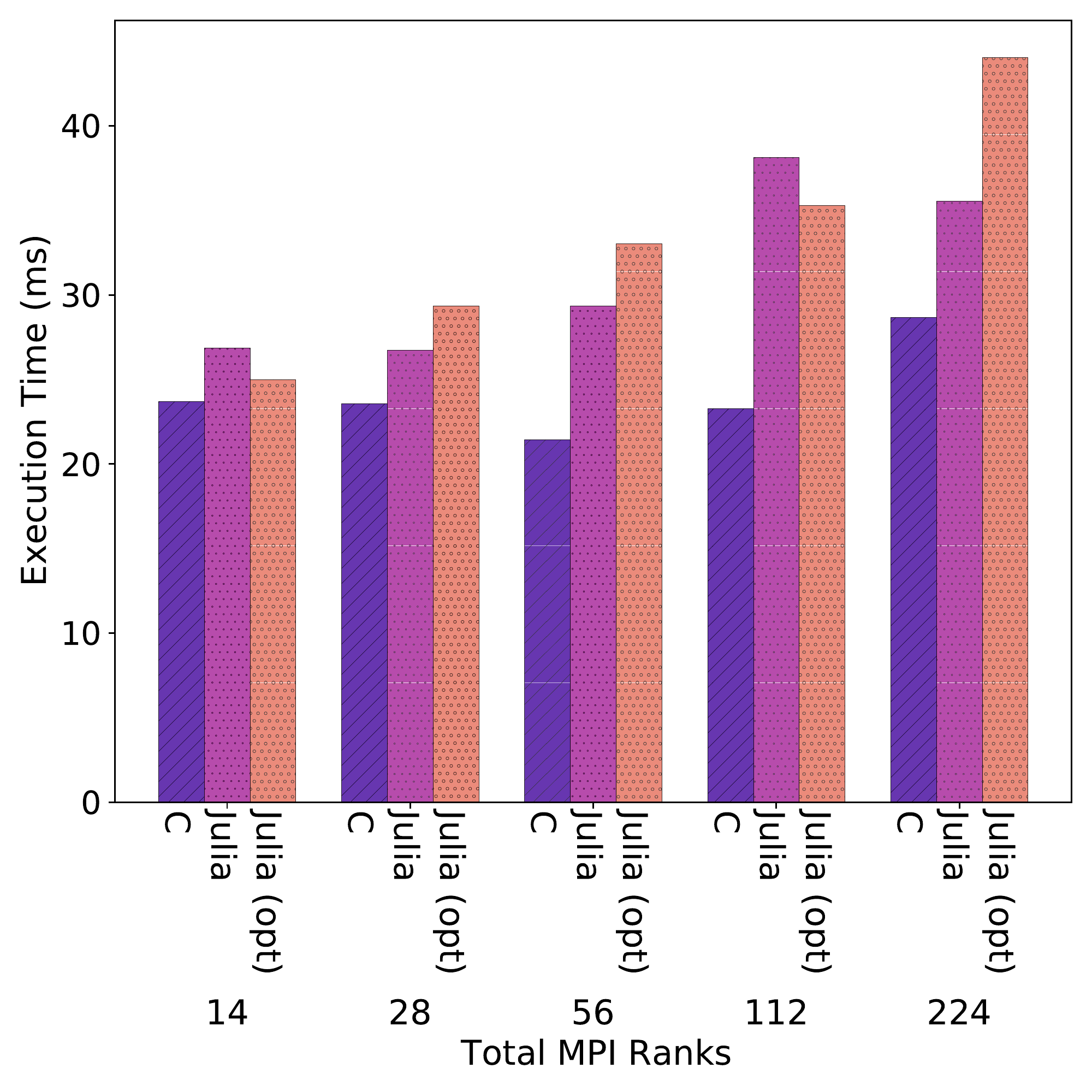}
        \label{fig:bsp-runtimes-100}
    }
    \caption{BSP benchmark total execution time.}
	\label{fig:bsp-runtimes}
\end{figure*}

Figure~\ref{fig:bsp-runtimes} shows mean execution times (10 communications in
Figure~\ref{fig:bsp-runtimes-10} and 100 in Figure~\ref{fig:bsp-runtimes-100}).
Note that Julia's MPI wrappers actually incur some overhead due from the
polymorphism in the \verb$MPI.jl$ implementation. We suspected this might have
a non-trivial overhead, so we also provided our own custom versions of the MPI
wrappers with static type signatures (denoted with ``opt'' in the figures) and
repeated experiments with those variants. This minor optimization essentially
eliminates string comparisons and runtime type coercion.  Communication in
Julia seems to be more expensive than in its lower-level counterpart. Our
optimization of bypassing dynamic dispatch by providing our own low-level
wrappers appears to have no noticeable effect on performance. Perhaps more
interestingly, we see that the performance difference scales with the number of
MPI ranks. Since we are actually increasing the {\em ranks per node} rather
than the absolute node count, the overheads in these experiments is likely
coming from effects {\em within} the node. This would be consistent with what
we saw in the spawn/fetch experiment.

\begin{figure*}
	\centering
    \subfigure[10 communication steps.]{
        \includegraphics[width=0.35\textwidth]{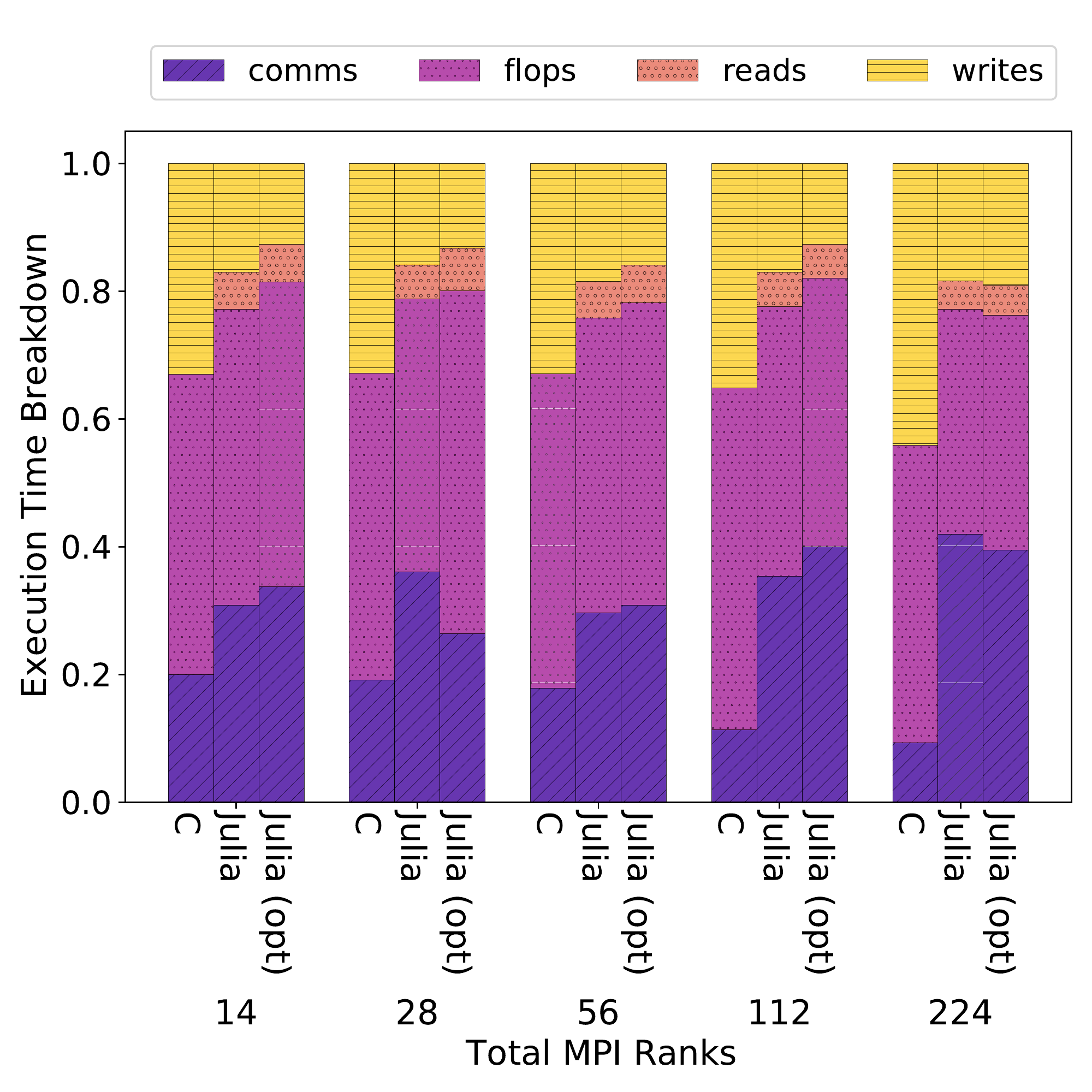}
        \label{fig:bsp-breakdown-10}
    }
    \subfigure[100 communication steps.]{
        \includegraphics[width=0.35\textwidth]{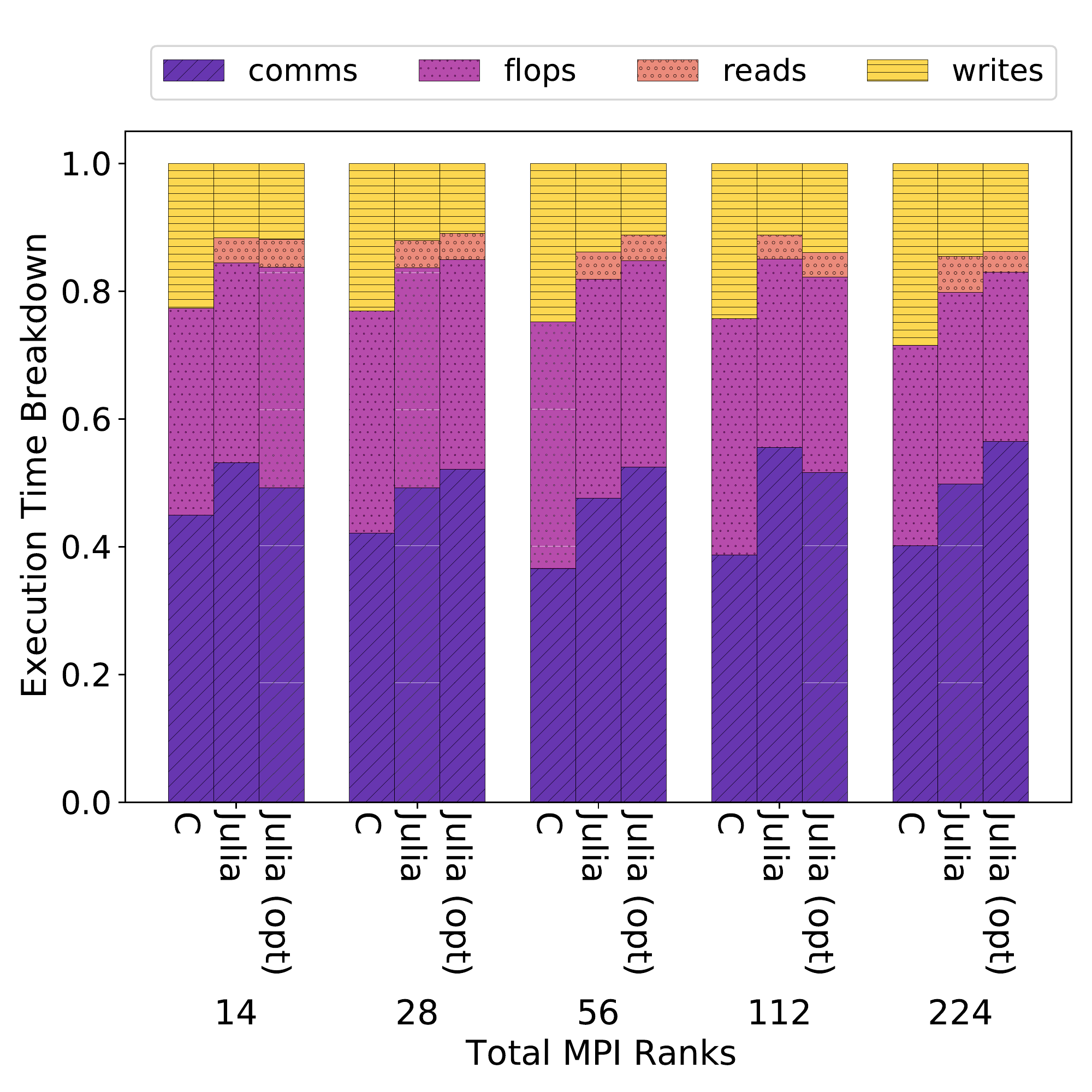}
        \label{fig:bsp-breakdown-100}
    }
    \caption{BSP experiment broken down per operation.}
    \label{fig:bsp-breakdown}
\end{figure*}

To verify that communication is the issue, we took a more detailed look at the
benchmark.  Figure~\ref{fig:bsp-breakdown} shows the same experiment, but with
the bars broken down per operation in the benchmark. In this figure the bars
show what percentage of execution each operation comprised. 
In both cases, reads are relatively cheap. Interestingly, the C version
writes (again, to an array) are more expensive relatively. Indeed, however, 
communications consume a more significant portion of the total execution time
in the Julia version. However, as the {\em number} of communications increases,
the difference between languages diminishes. Because both versions are ultimately
using the same MPI implementation, we can rule out the communications themselves.
While we are still investigating this effect, we suspect the culprit might be
how Julia manages the underlying workers. Note that the figures here exaggerate
the differences since the total execution time of this benchmark is in the tens
of milliseconds.

%C/MPI and Julia/MPI experiments showing that floating point operations are
%least expensive and write operations are most expensive. Results from
%Figure~\ref{fig:bsp-runtimes} confirm that communication operations in
%Julia/MPI incur some overheads as compared to communication operations deployed
%in C/MPI. MPI in Julia is connected in an all-to-all (not master-slave) format
%but in a "lazy" way wherin master process will be able to connect with all
%workers but a worker process will only establish a connection with another
%worker process if a remote invocation between them exists. Communciation
%latencies increase with added number of processes especially for Julia/MPI,
%because each process makes an attempt to communicate with another off machine
%process over the network. \verb|MPI.jl| is simply a Julia wrapper for the MPI
%protocol. It ccalls the C library functions and the latency incorporated in
%Julia/MPI communications can be attributed to \verb|ccall| on sufficiently big
%array structures instead of primitive data types like \verb Int64, \verb.Float64.  
%etc. Moreover, for each \verb|MPI.Send()| function call that we use in
%the experiment, Julia makes at least two function calls before it ccalls
%\verb|MPI_SEND| operation. These overheads add up for 10000  send and recieve
%operations that we call on increasing number of processes.

\begin{figure}
	\centering
	\includegraphics[width=\columnwidth]{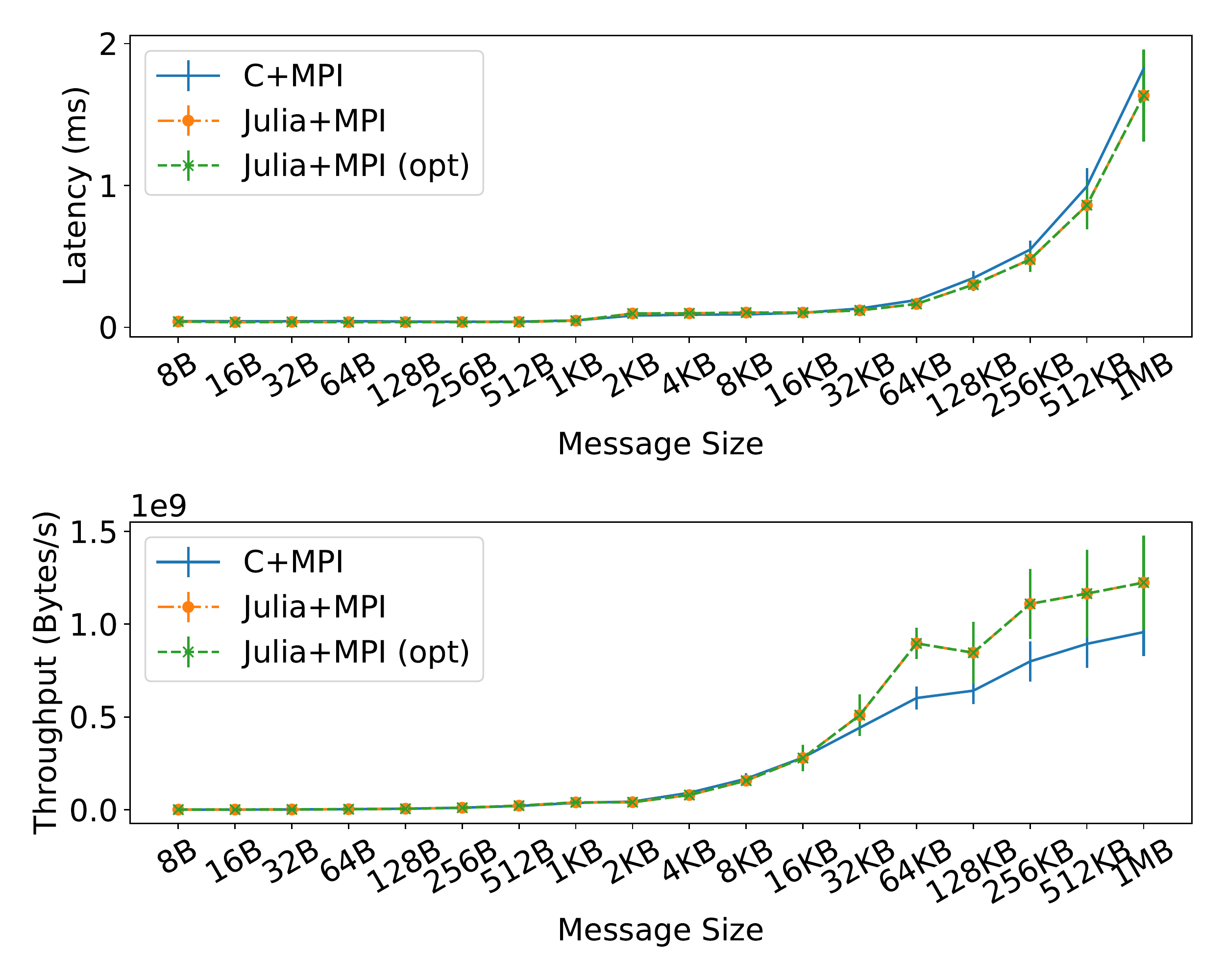} 
	\caption{ Latencies (above) and throughput (below) for an MPI ping-pong experiment 
    with both Julia and C implementations.}
	\label{fig:pp}
\end{figure}

To further verify that the communications themselves were not the culprit, we
conducted a ping-pong experiment with MPI for both languages, using both the
normal Julia MPI wrappers and our custom implementation. The results are shown
in Figure~\ref{fig:pp}. We measure the ping-pong latency of a round-trip send
between two remote MPI ranks and report the mean latency (top) as we vary
message size (up to 1MB). We also report the harmonic mean of message
throughput (bottom). Outliers have been removed as in our spawn/fetch experiment.
In this case, we see that Julia actually {\em outperforms}
its lower-level counterpart, though we do see overlap in the error bars, and
differences here will be in the $\mu$s range. Again, our minor optimization for
Julia's MPI wrappers appears to have no distinguishable effect.  While the
results are at odds with what we saw in our BSP experiment, we note that an
obvious difference in these scenarios is that in this case, we only have two
Julia worker processes, one per node. The improved communication performance in
this scenario further supports our hypothesis above about quirks in worker
management.

\subsection{HPCG}
\label{sec:hpcg}

The microbenchmarks give us a preliminary idea of how various native Julia
constructs and operatives perform, and what kind of distributed performance we can
expect to achieve. However, it is helpful to have a more realistic application to
perform an evaluation with. Our goal here was to choose an application that is
widely used and represents a broad range of HPC applications. To that end, we chose
a standard, state-of-the-art benchmarking application that is used to evaluate
top supercomputers today. We also wanted our effort to be useful for others to investigate
Julia's large-scale performance (perhaps on a much larger system).

The High-Performance Conjugate Gradients (HPCG)
benchmark~\cite{DONGARRA:HPCG:2013, Dongarra2015HPCG} fits that bill.  It is
a tool for ranking computer systems based on a simple additive Schwarz,
symmetric Gauss-Seidel preconditioned conjugate-gradient solver. It constructs
a 3-dimensional, 27-point stencil matrix such that each point, represented by
coordinates $i$, $j$, and $k$ depends on the value of the surrounding 26
points.

The generated matrix ($A$) is sparse and has at least 27 non-zero values for
interior points~\cite{Dongarra2013HPCGSpec}. The boundaries have 7 to 18
non-zero points depending on the boundary position. The Matrix $A$ is symmetric
with positive eigenvalues (symmetric positive definite). It is stored in
Compressed Sparse Row (CSR) format.  The global domain of $M_x$, $M_y$, and
$M_z$ in the $x$, $y$, and $z$ dimensions is distributed into a local domain of
$N_x$, $N_y$, and $N_z$. Each local sub-grid is assigned to a process (MPI
rank). The total number of processes is provided as an input parameter.  These
processes are distributed into a 3D domain, where the total number of processes
$P = P_x \times P_y \times P_z$. The global domain $M_x$, $M_y$, $M_z$ can be
calculated as $(P_x \times N_x)(P_y \times N_y)(P_z \times N_z)$. A restriction
is also imposed on the domain for accuracy and efficiency purposes. Minimum
grid size in each dimension must be at least
16 and the grid should be a multiple of 8.

\begin{figure*}
	\centering
	\subfigure[$N_x = N_y = N_z = 16$]{
        \includegraphics[width=.30\textwidth]{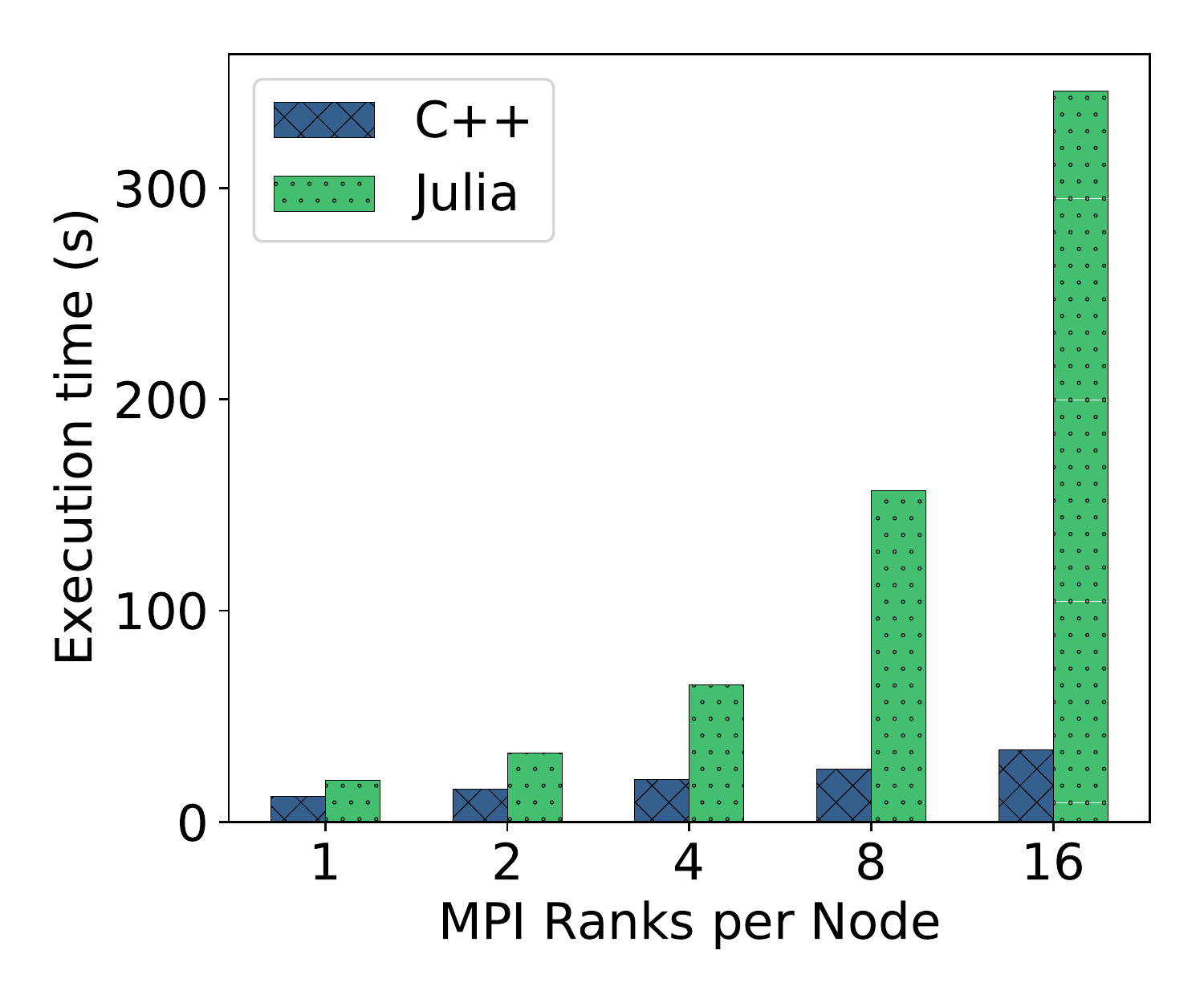}
        \label{fig:HPCG_EXP-16}
	}
	\subfigure[$N_x = N_y = N_z = 32$]{
        \includegraphics[width=.30\textwidth]{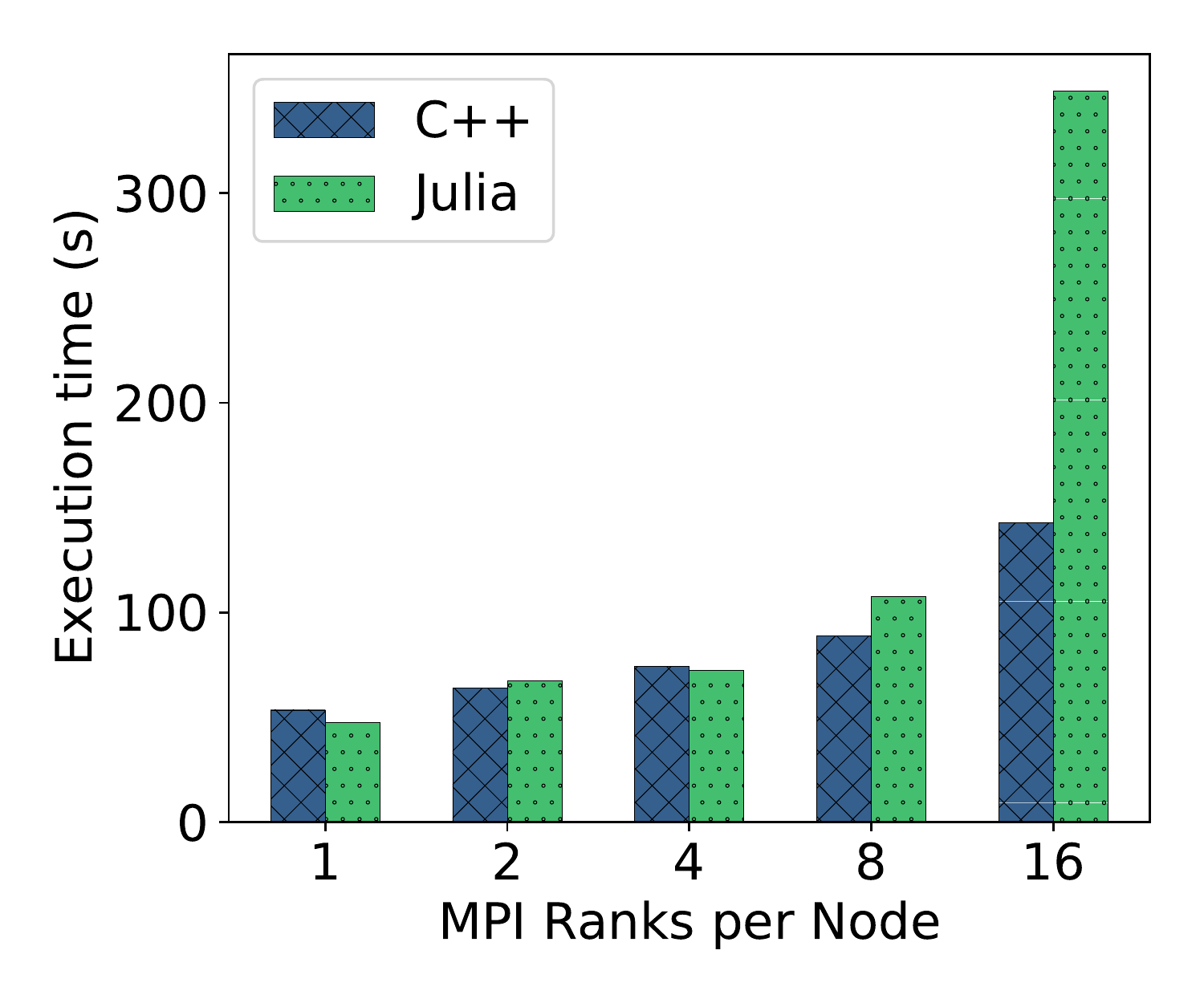}
        \label{fig:HPCG_EXP-32}
	}
	\subfigure[$N_x = N_y = N_z = 64$]{
        \includegraphics[width=.30\textwidth]{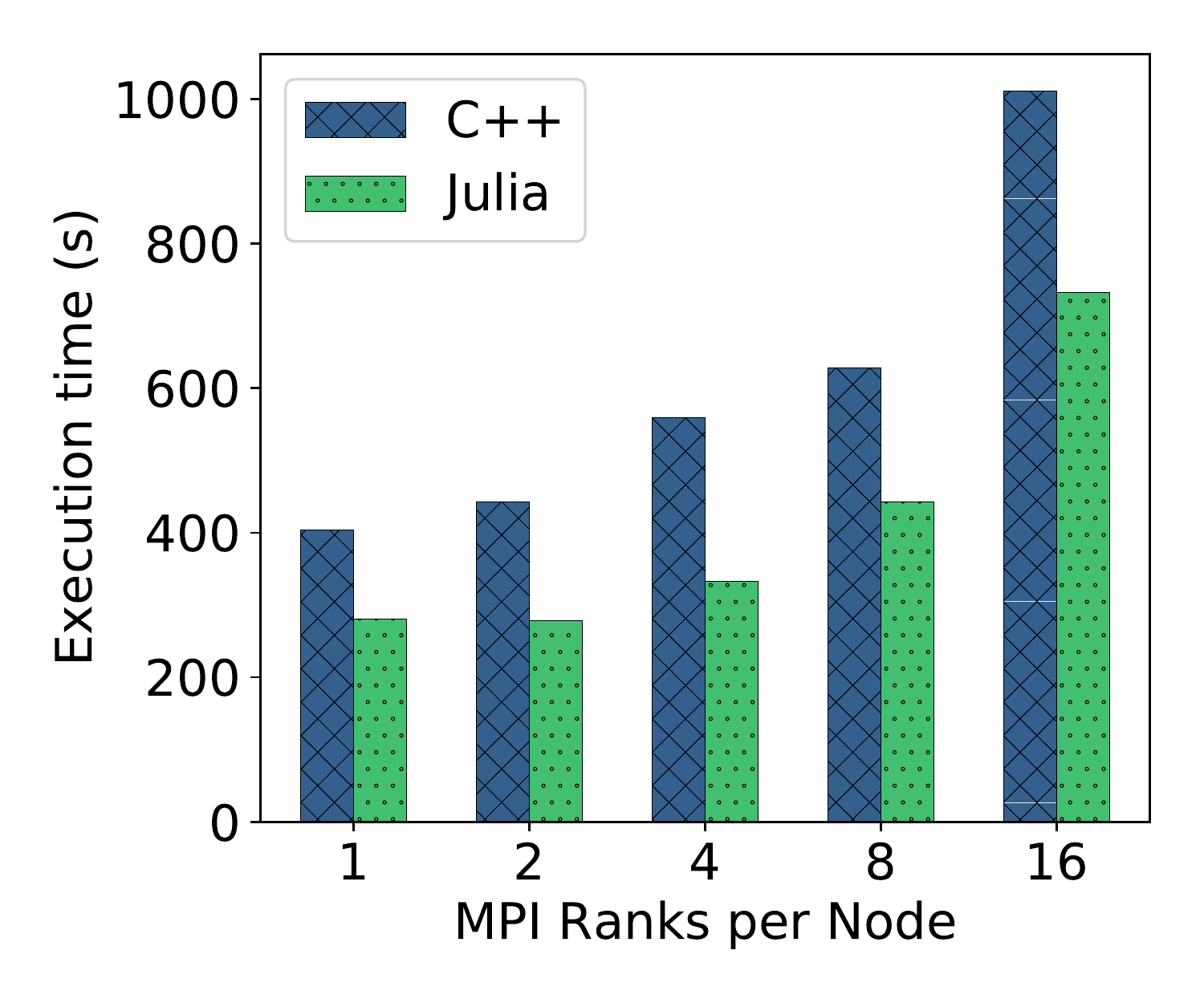}
        \label{fig:HPCG_EXP-64}
	} 
    \caption{Strong scaling of the standard C++ implementation of HPCG compared to our Julia port.}
	\label{fig:HPCG_EXP}
	\end{figure*}

We ported HPCG version 3.1 \footnote{Available at
\url{https://hpcg-benchmark.org}} to Julia and ran it varying the local domain
size ($N_x = N_y = N_z = 16,32,64$) and varying the number of MPI ranks per
node (from 1 to 16).  The results shown in Figure~\ref{fig:HPCG_EXP} are single
iteration runs. Each
subfigure has a different local domain size, so each is showing strong scaling
for a particular parameter. We compare it to the standard C++/MPI version of
the benchmark (without OpenMP support).

The results here are consistent with the results from our BSP experiment. With
a small local domain size (see Figure~\ref{fig:HPCG_EXP-16}), computation
comprises a relatively small portion of the execution time, as each processor
will have little work to do before having to communicate. In this setting we
will see the Julia overheads that we have seen in previous experiments. The
remarkable thing to note here is that as the problem size per rank grows (see
Figure~\ref{fig:HPCG_EXP-64}), thus increasing the compute/communication ratio,
Julia actually starts to outperform the C++ version of the benchmark by as much
as 20\%.  Julia's scaling also improves (as the C++ version worsens) as the
number of ranks per node is increased for the large local domain size.

Note that this benchmark communicates primarily with point-to-point
operations (\verb.MPI_Send.  and \verb.MPI_Recv. in the halo exchange). Though
there are uses of collectives and barriers scattered throughout the code (e.g.
\verb.MPI_AllReduce.), they comprise very little of the total execution time.

The results from HPCG clearly suggest that Julia has significant potential,
even for HPC applications that heavily employ synchronous communication. We
note that our port of HPCG is currently a prototype, and thus we expect its
performance relative to its low-level counterpart will improve further as we
adopt more idiomatic usage of the language and as the language itself improves.

\section{Discussion}
\label{sec:discuss}

We now outline some of the issues we encountered with porting a large benchmark
from a low-level systems language to a Julia, a high-level managed language.
In order to have a fair comparison between C++ and Julia's version of HPCG it
was essential that we adopt it within the confines of Julia's abstractions.  One
major difference between C++ and Julia is obviously the use of
managed memory, which made porting HPCG's pointer-heavy manipulation of
sparse matrices a challenge.

Structures in HPCG like \verb.matrixValues., \verb.mtxIndG., \verb.mtxIndL.,
which are pointer to pointer structures, are ported to Julia using 2D arrays.
\verb.matrixdDiagonal. is an array of pointers that refer to the addresses of
diagonal elements in \verb$matrixValues$. In Julia, this is replaced by an array
called \verb.curcols. which stores the column indices of \verb.matrixValues.
where the diagonal elements are stored. Diagonal elements can be fetched for
a given row index by accessing \verb.curcols. on the same index.  Pointer
arithmetic is substituted by careful indexing as Julia is 1-indexed (\'{a} la
Fortran) and C++ is 0-indexed.  
%An example of how Julia's abstractions come
%into play is when \verb.Vector., a small structure and its functions like
%ScaleVectorValue(), ZeroVector(), DeleteVector() is ported easily without
%creating a seperate structure and removes the overhead of including another
%file into the main code.

In the function \verb.FillRandomVector()., the C++ version uses \verb.rand().
from the C standard library to fill a vector with pseudo-random values. Because
Julia uses a different PRNG routine, and because we had to compare outputs
between versions for correctness, we modified this routine to fill vectors with
known, fixed values.

In the procedure \verb.ComputeOptimalShapeXYZ. in the C++ version, a complex
\verb.for. loop uses \verb.MixedBaseCounter. structures and its corresponding
functions to initialize, check bounds, and increment the
\verb.MixedBaseCounter. structure.  These are ported using Julia's looping
constructs, which are slightly different.  A similar approach is used in
\verb$GenerateProblem_ref$.

Primitive data types in Julia passed as parameters to a function and mutated in
the function body will not retain their altered values on return (as is
standard in high-level languages). However, HPCG uses a common C/C++ idiom of
passing in pointers to primitive types to have the values mutated by the
callee. The Julia versions of such functions must have slightly different
structure to account for this. When possible, we followed Julia's documented
best
practices\footnote{\url{https://docs.julialang.org/en/v1/manual/performance-tips/index.html}}
for performance programming. This included making code type-stable, eliminating
use of any global variables,  eliding bounds checks for known-length arrays
(\verb.@inbounds.), and carefully tuning core HPCG math kernels for performance
bottlenecks, e.g. by instrumenting the garbage collector and using Julia's
native profiler. We still, however, believe there is room to improve the
performance. 

While not reported in the figures, we did see a notable difference in the
variance of Julia's runs on the larger code. This is not entirely surprising,
as jitter is a known pain point for applying managed languages in HPC settings
(this is, of course, true for the underlying system as
well~\cite{HOEFLER:2010:NOISE, FERREIRA:2008:OSNOISE, MORARI:2011:OSNOISE}).
In Julia's case, two primary sources of non-determinism include the garbage
collector and the JIT compiler. While others have looked at these effects for
other languages~\cite{AKERET:2015:HOPE, PRABHU:2019:MOYA}, we believe there is
opportunity to optimize these features to avoid jitter for Julia (and other
managed, technical computing languages) in synchronous HPC settings.

\section{Related work}
\label{sec:related}

There has long been interest in bringing the convenience of high-level
languages to bear for HPC.  High-performance Fortran~\cite{MERLIN:1995:HPF},
Chapel~\cite{CHAMBERLAIN:2007:CHAPEL}, X10~\cite{CHARLES:2005:X10}, and
Erlang~\cite{JOHANSSON:2000:HPE}, and Regent~\cite{SLAUGHTER:2015:REGENT} are
notable examples. While Python was not designed with HPC computing in mind,
there are efforts in that community to adapt it for high-performance
parallelism~\cite{Ross2016MPI4Py,Gomez2015HPCPython,
Hand2017Nbodykit,Paraskevakos2018Molecular}.

Most recent work on Julia focuses on adapting it for practical problems in
technical computing, several of which we described in Section~\ref{sec:julia}.
From the HPC perspective, the most striking example is
Celeste~\cite{Regier2015CelesteVI}, the first large-scale HPC application
written entirely in Julia. It was reported to achieve petascale performance on
a NERSC supercomputer.  However, Celeste is not structured to be
communication-intensive, unlike many HPC applications, so its success, while
remarkable, does not yet demonstrate Julia's broad applicability for
large-scale, communication-intensive HPC applications. 

Others have evaluated Julia in a multi-core
setting~\cite{Knopp2014JuliaMultiThreading, bean2013ManyCoreJulia}. Hunold and Steiner
study Julia's MPI bindings~\cite{HUNOLD:2020:JULIA}, though in a more focused setting. Our results with the more
substantial HPCG benchmark, however, do line up with their findings. 

There also has been extensive work done in programming GPUs with Julia.
In addition to the the foundational GPUArrays package JuliaGPU\footnote{Available at
	\url{https://juliagpu.org/}}   is a package in Julia
which encompasses support for both NVIDIA and AMD GPUs. The NVIDIA CUDA 
boasts mature and full-featured packages for both low-level kernel programming as well as working with high-level operations on arrays. All versions of Julia are supported, and the functionality is actively used by a variety of applications and libraries.

\section{Conclusions and future work}
\label{sec:conc}

In this paper, we evaluated Julia using a range of microbenchmarks and
macrobenchmarks both for shared-memory parallelism within the node and for
distributed parallelism in a cluster computing setting. We showed that while
Julia has slight overheads for communication between processes, these overheads
are often outweighed by gains over its lower-level counterparts (C/C++). We
demonstrated using a custom synthetic BSP benchmark and the first ever Julia
port of a large-scale, HPC benchmark application (HPCG) that Julia can
outperform C++/MPI even in a multi-node setting with synchronous communication.
We believe these findings open up several lines of investigation into Julia as
a first-class HPC language. First, we plan to extend our experiments to add
comparisons to LLVM-compiled C/C++ code. We also plan to further investigate
the source of communication anomalies between Julia's worker processes.
Finally, we hope to explore optimizations for sources of non-determinism in the
Julia runtime and to scale our experiments to hundreds or even thousands of
nodes.

\section*{Acknowledgements}
This work is made possible with support from the United States National Science
Foundation (NSF) via awards CNS-1718252 and CNS-1730689.

\balance
\bibliographystyle{IEEEtran}
\bibliography{amal,kyle}
\end{document}